\documentclass[sigconf]{acmart}

\AtBeginDocument{%
  }

\setcopyright{none}
\acmDOI{}
\acmISBN{}

\acmConference[LOCO '24]{1st International Workshop on Low Carbon Computing}{December 3, 2024}{Glasgow, Scotland, UK}

\begin{document}

\title{Green Metrics Tool: Measuring for fun and profit}


\author{Geerd-Dietger Hoffmann}
\affiliation{%
  \institution{Green Coding Solutions}
  \city{Berlin}
  \country{Germany}}
\email{didi@green-coding.io}

\author{Verena Majuntke}
\affiliation{%
  \institution{HTW}
  \city{Berlin}
  \country{Germany}
}
\email{verena.majuntke@htw-berlin.de}

\begin{abstract}
The environmental impact of software is gaining increasing attention as the demand for computational resources continues to rise. In order to optimize software resource consumption and reduce carbon emissions, measuring and evaluating software is a first essential step. In this paper we discuss what metrics are important for fact base decision making. We introduce the \emph{Green Metrics Tool} (GMT), a novel framework for accurately measuring the resource consumption of software. The tool provides a containerized, controlled, and reproducible life cycle-based approach, assessing the resource use of software during key phases. Finally, we discuss GMT features like visualization, comparability and rule- and LLM-based optimisations highlighting its potential to guide developers and researchers in reducing the environmental impact of their software.
\end{abstract}


\maketitle

\section{Introduction}
The significance of the environmental impact of software and hardware has grown in recent years \cite{9407142}. With the increasing demand for computational resources \cite{wef2024}, the reduction of such and associated carbon emissions becomes progressively important. Currently 8–10\% of total electricity production \cite{10.1145/3613207} is used by the information and communication infrastructure. With new AI technology this is projected to further increase \cite{BERTHELOT2024707}.

In order to optimize software, the measurement and evaluation of resource usage is the first essential step. In this paper we (1) discuss metrics which are required to accurately assess the resource usage of software. (2) We present a novel framework, the \emph{Green Metrics Tool}, specifically designed to facilitate the precise measurement of metrics across different operating systems and along the entire software life cycle such as installation, runtime, and removal. (3) Lastly, we discuss the visualization of data and optimization recommendations the GMT provides to assess and improve resource usage. All tools and systems are published open source\cite{greenmetrics} under the GNU Affero General Public License\cite{agpl3}.

\section{Related Work}

The measurement of resource consumption and performance has been an area of active research since the inventions of computers. Early approaches primarily focused on hardware-level measurements, utilizing specialized equipment to monitor energy usage in real-time \cite{bellosa2001}\cite{Tiwari1996}. However, these methods were often impractical for widespread adoption due to the need for expensive hardware and the complexity of setup \cite{Abdurachmanov_2015}\cite{srikantaiah2008energy}. Other tools like Greenframe.io \cite{greenframe}, Kepler \cite{sustainablecomputing}, Scaphandre \cite{scaphandre} are geared towards assessing the general energy consumption but do not focus on fine grained benchmarks.

\section{Challenges and Metrics}\label{sec:metrics}

The accurate measurement of software resource consumption is a complex task due to the inherent variability in system behavior and the information required to assess environmental impact. The precision of any measurement is highly dependent on the system environment, i.e., hardware specifications, operating system configuration, and background processes. While CPU energy consumption is often the primary measurement focus - likely due to the availability of tools such as Intel’s Running Average Power Limit (RAPL) - this metric alone is insufficient for a holistic assessment as not all problems are CPU bound. In our analysis we identified $15$ metrics, e.g., thermal characteristics, disk/network activity, memory utilization, and execution time to name a few. Each of these parameters contributes to a fuller understanding of the environmental impact of software. Moreover, certain system-level inefficiencies, such as resource overprovisioning, have to be considered. Accurate measurements are essential for distinguishing the effects of code edits, as inaccuracies can lead to incorrect conclusions and misguided optimization efforts. 

\section{System Architecture}
\begin{figure}[h]
  \centering
  \caption{The GMT Architecture}
  \includegraphics[width=220px]{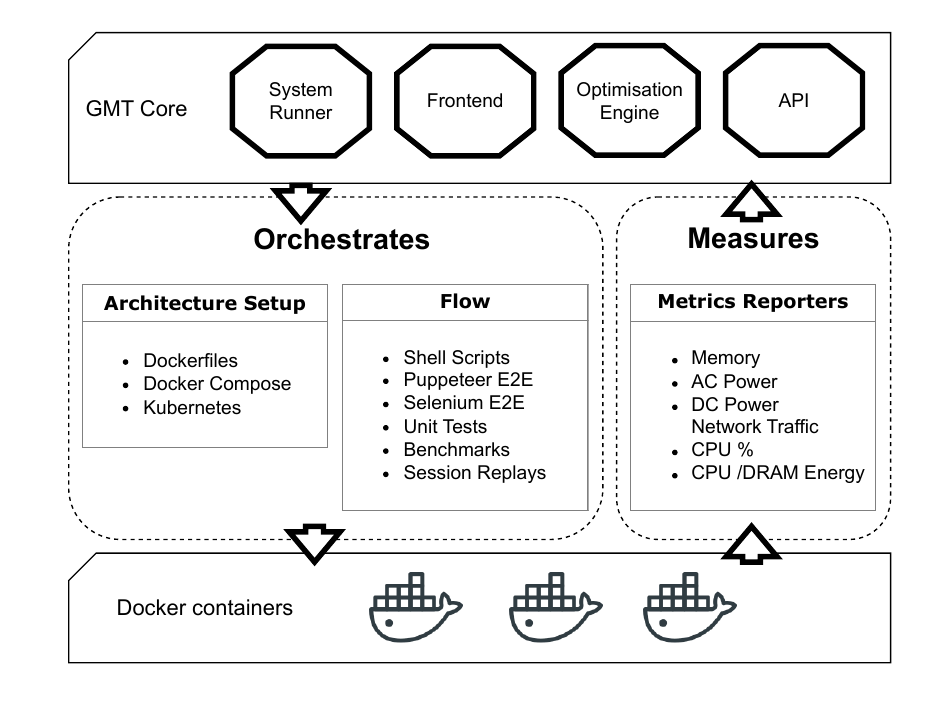}
  \vspace{-20pt}
  \label{fig:arch}
\end{figure}

The core objective of the GMT is to enable highly accurate reproducible measurements. The architecture overview is shown in Figure \ref{fig:arch}. To achieve reproducibility and ensure a controlled and isolated environment, the tool employs containerization technology (Docker \cite{Docker}) to orchestrate the measurement environment. By using containers, the GMT standardizes the testing scenario, reducing the variability introduced by different system configurations and environments enabling precise measurements for all parts of the system, including network traffic. A configuration is used, modeling how the tool is deployed, shown as \emph{Architecture Setup}. Furthermore, the specification of a \emph{usage scenario} modeling how the system is used (shown as \emph{Flow}) is required. Since this is also configured in a container, it can be tailored to whatever is most suitable for the software being benchmarked like \emph{curl}\cite{Stenberg2023curl}, \emph{Puppeteer}\cite{Puppeteer2023} or \emph{Playwright}\cite{Playwright2023}. The tool also offers an interactive mode which is meant for logging resource consumption in deployed systems. This is used to benchmark running AI systems, for example\cite{hoffmann2024improving}. The tool also includes a \emph{web fronted} and an \emph{optimisation engine} which are described in section~\ref{sec:data}. Further there is a comprehensive \emph{API} to extract the benchmark data.

The GMT utilizes a range of \emph{Metrics Reporters} which are small, specialized, configurable programs that log performance metrics depending on the machine the benchmark is executed on. These reporters collect data on energy consumption, CPU utilization, memory usage, various temperature sensors and other relevant parameters, providing a comprehensive view of the software's environmental impact. Notably, the GMT is designed to minimize its own interference during the benchmarking process. While the benchmark is running, the tool refrains from performing any processing, writing the collected data directly to a file. This design choice was empirically validated to impose minimal overhead ($<1\%$), ensuring that the measurement process itself does not skew the results.

To support sharing specialized measurement hardware, large-scale periodic benchmarking and avoiding the pitfalls of consumer hardware, the Green Metrics Tool offers a cluster deployment option. This enables benchmarking jobs to be run on certain signals like a git commit or periodic measurements of resource consumption over the development time. It also enables the use of expensive specialized hardware, required for precise measurements, as multiple people can share the same physical machine. 

\subsection{System Calibration and Pre-Measurement Checks}
To further enhance measurement accuracy, the GMT performs system checks before initiating the benchmarking process. One critical factor considered is the CPU temperature, as fluctuations in temperature can significantly impact energy consumption readings. It is also important to account for temperature changes over time as this overhead needs to be accounted for through cooling. Additionally, the GMT can disable CPU features, such as Turbo Boost and dynamic frequency scaling as they can introduce variability in performance metrics. Another critical aspect of the GMT's methodology is a calibration script which measures the baseline resource utilization and temperature of the system in its idle state, establishing a reference point for subsequent measurements.

\subsection{NOP Linux and Interrupt Reduction}
To mitigate the impact of operating system-induced variability on measurements, the GMT integrates with NOP Linux \cite{nopLinux2024}, a specialized Linux flavour designed to minimize system OS activity. Operating system interrupts can disrupt the measurement process by consuming CPU cycles. This alters energy consumption patterns as most measurement devices report on a per core or whole machine level. NOP Linux disables these services providing a more stable environment for the measurements while keeping the system as close to an off the shelf system as possible. 


\subsection{Lifecycle Assessment and Comprehensive Measurement}
The GMT's approach to software measurement encompasses the entire software life cycle, from initial installation to eventual removal. This holistic view allows developers to make data-driven decisions that reduce the environmental impact of their software along the life cycle. The life cycle stages considered by the GMT include \textbf{Baseline}, which is the measurement of the system's idle state to establish a reference point; \textbf{Installation}, involving the evaluation of the resource utilization during the software installation and build process; \textbf{Boot}, assessing resource characteristics during the software's startup phase; \textbf{Idle}, measuring the software's behavior when it is running but not actively being used; \textbf{Runtime}, analyzing the software's resource consumption during active use; and \textbf{Removal}, evaluating the impact of the software's uninstallation on the system.

\section{Data Visualization and Optimization Recommendation}
\label{sec:data}
The GMT includes an interface for data visualization and comparison allowing users to explore collected data in detail, facilitating comparisons between different versions or configurations. The comparison view is particularly useful for identifying changes that may have led to resource improvements or regressions. By providing a clear and intuitive visualization, the GMT empowers developers to make informed decisions about how to optimize their software.

The GMT also offers optimization recommendations based on the collected data. These recommendations are generated after the analysis of the software across the different life cycle stages. For this purpose, a rule based system looks at the values and flags abnormalities like over provisioning of resources, long boot times, low IPC counts, high page faults, etc. A configurable external LLM (Llama\cite{Llama2023}, ChatGPT\cite{OpenAIChatGPT2023}, Mistral\cite{MistralAI2023}, etc.) is prompted for improvement recommendations for code segments which have shown to have a high resource usage. In this step the code segment is copied and a prompt is created queering the LLM to try to "improve" this segment. There is also the option for an LLM to "rate" code segments and suggest improvements. This is done by prompting the LLM with the respective code segments.

\section{Conclusion and Future Work}
The Green Metrics Tool represents a significant advancement in the field of sustainable software development. By providing precise, reliable measurements and minimizing interference during the benchmarking process, the GMT enables developers to optimize their software based on accurate and comprehensive data. The integration of containerization, NOP Linux, and extensive pre-measurement checks ensures that the tool can deliver consistent and reproducible results. Moreover, the GMT's lifecycle-based approach to software assessment and its robust data visualization capabilities make it an invaluable resource for developers seeking to reduce the environmental impact of their software. Future work includes extending the GMT to have more recommendations, more reporters and finer grain analytics. As the demand for sustainable software continues to grow, tools like the GMT will play a crucial role in helping developers meet these challenges. 

\bibliographystyle{ACM-Reference-Format}
\bibliography{loco}

\newpage

\end{document}